\RequirePackage{fix-cm}
\RequirePackage{rotating}
\documentclass[runningheads]{llncs}
\usepackage{graphicx,booktabs,xcolor}
\usepackage{amsfonts}
\usepackage{amsmath}

\def\BibTeX{{\rm B\kern-.05em{\sc i\kern-.025em b}\kern-.08em
    T\kern-.1667em\lower.7ex\hbox{E}\kern-.125emX} }

\usepackage[outdir=./]{epstopdf}
\begin{document}
\title{Cellular Traffic Prediction and Classification: a comparative evaluation of LSTM and ARIMA}
\author{Amin Azari\inst{1} \and
Panagiotis Papapetrou\inst{1} \and
Stojan Denic\inst{2} \and
Gunnar Peters\inst{2}
}
\authorrunning{Azari et al.}
\institute{Dept. of Computer and Systems Sciences \\ Stockholm University, Sweden
\email{\{amin.azari,panagiotis\}@dsv.su.se}
\and
Huawei, Sweden
\email{\{stojan.denic,gunnar.peters\}@huawei.com}
}
\maketitle           
\begin{abstract}
Prediction of user traffic in cellular networks has attracted profound attention for improving resource utilization. In this paper, we study the problem of network traffic traffic prediction and classification by  employing standard machine learning and statistical learning time series prediction methods, including long short-term memory (LSTM) and autoregressive integrated moving average (ARIMA), respectively. We present an extensive experimental evaluation of the designed tools over a real network traffic dataset. Within this analysis, we explore the impact of different parameters to the effectiveness of the predictions. We further extend our analysis to the problem of network traffic classification and prediction of traffic bursts. The results, on the one hand, demonstrate superior performance of LSTM over ARIMA in general, especially when the length of the training time series is high enough, and it is augmented by a \textit{wisely}-selected set of features. On the other hand, the results shed light on the circumstances in which, ARIMA performs close to the optimal with lower complexity.

\end{abstract}

\section{Introduction}\label{sec:introduction}

A major driver for the beyond fifth generation (5G) wireless networks consists in offering the wide set of cellular services in an energy and cost efficient way \cite{Sad6g}. Toward this end, the legacy design approach, in which resource provisioning and operation control are performed based on the peak traffic scenarios, are substituted with predictive analysis of mobile network traffic and proactive network resource management \cite{RaUrllc,Db6g,Sad6g}. Indeed, in cellular networks with limited and highly expensive time-frequency radio resources, precise prediction of user traffic arrival can contribute significantly in improving the resource utilization \cite{RaUrllc}. As a result, in recent years, there has been an increasing interest in leveraging machine learning tools in analyzing the aggregated traffic served in a service area for optimizing the operation of the network \cite{RnnScale,DrxSelf,DrlBsSleep,UavML}.  Scaling of fronthaul and backhaul resources for 5G networks has been investigated in \cite{RnnScale} by leveraging methods from recurrent neural networks (RNNs) for traffic estimation. Analysis of cellular traffic for finding  anomaly in the performance and provisioning of on-demand resources for compensating such anomalies have been investigated  in \cite{UavML}. Furthermore,  prediction of light-traffic periods, and saving energy for access points (APs) through sleeping them in the respective periods has been investigated in \cite{DrxSelf,DrlBsSleep}. Moreover, Light-weight reinforcement learning for figuring out statistics of interfering packet arrival over different wireless channels has been recently explored \cite{SelfMl}. While one observes that analysis of the aggregated traffic at the network side is an established field, there is lack of research on the analysis and understanding at the user level, i.e., of the specific users' traffic arrival. In 5G-and-beyond networks, the (i) explosively growing demand for radio access,  (ii) intention for serving battery- and  radio-limited devices requiring low-cost energy efficient service  \cite{SelfMl},  and  (iii) intention for supporting ultra-reliable low-latency communications \cite{RaUrllc}, mandate studying not only the aggregated traffic arrival from users, but also studying the features of traffic arrival in all users, or at least for critical users. A critical user could be defined as a user whose quality-of-service (QoS) is at risk due to the traffic behavior of other devices, or its behavior affects the QoS of other users.  Let us exemplify this challenge in the sequel in the context of cellular networks.
 \begin{figure}[!tb]
        \centering
                \includegraphics[width=\columnwidth]{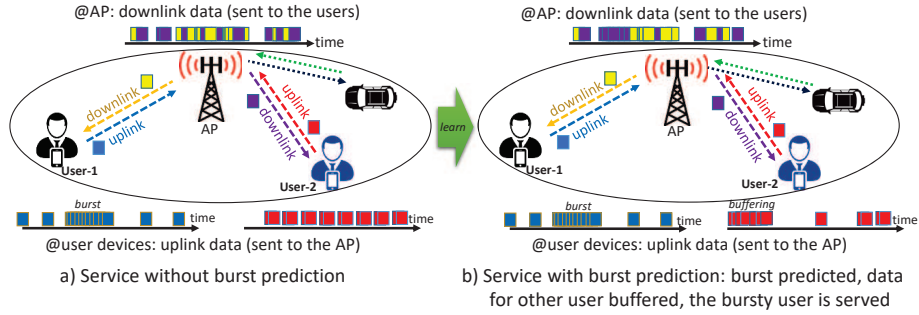}
                \caption{A communication network including access point, users, and uplink and downlink data communications. a) service is offered without prediction of bursts, b) service is adapted to the probability of occurrence of bursts.} 
                \label{sys}
\vspace{-6mm}
\end{figure} 

\smallskip
\noindent \textbf{Example.} Fig. ~\ref{sys}~(a) represents a communication network in which, an AP is serving users in the uplink (towards AP) and downlink (towards users). One further observes that traffic from user-2 represents a semi-stable shape, which is usually the case in video streaming, while the traffic from user-1 represents a bursty shape, which could be the case in surfing and on-demand file download. One observes that once a burst in traffic of user-1 occurs, the server (i.e. AP) will have difficulty in serving both users in a timely manner, and hence, QoS degradation occurs. Fig. ~\ref{sys}~(a) represents a similar network in which, AP predicts the arrival of burst to user-1, immediately fills the buffer of user-2. Thus, at the time of arrival of burst for user-1, user-2 will require minimal data transfer from the AP, and hence, QoS degradation for user-2 will be prevented. Backed to this motivation,  the remainder of this paper is dedicated to investigating the feasibility of exploiting the traffic history at the user level and employing it for future traffic prediction via machine learning and statistical learning approaches.

\smallskip
\noindent \textbf{Research problem.} 
Let us assume time in our problem is quantized into intervals of length $\tau$ seconds. The research problem tackled in this work could be stated as follows:
\textit{Given the history of traffic arrival for a certain number of time intervals, how accurately can we estimate (a) the intensity of traffic in the next time intervals, (b) the occurrence of burst in future time intervals (c) the application which is generating the traffic?} 

This problem can be approached as a time series forecasting problem, where for example, the number of packet arrivals in each unit of time constitutes the value of the time series at that point. While the literature on time series forecasting using statistical and machine learning approaches is mature, e.g., refer to \cite{TiSerNn,Hyb} and references herein, finding patterns in the cellular traffic and making the decision based on such prediction is never an easy task due to the following reasons \cite{CellTrChar}. First, the traffic per device originates from different applications, e.g. surfing, video and audio calling, video streaming, gaming,  and etc. Each of these applications could be mixed with another, and could have different modes, making the time series seasonal and mode switching. Second, each application can generate data at least in two modes, in active use and in the background, e.g. for update and synchronization purposes. Third, each user could be in different modes in different hours, days, and months, e.g. the traffic  behavior in working days differs  significantly from the one in the weekends. Forth, and finally, the features in the traffic, e.g., the inter-arrival time of packets, vary significantly in traffic -generating applications and activity modes. 
  
\smallskip
\noindent \textbf{Contributions.}
Our contributions in this paper are summarized as follows:
\begin{itemize}
    \item We present a comprehensive comparative evaluation for prediction and classification of network traffic; autoregressive integrated moving average (ARIMA) against the long short-term memory (LSTM);
    \item we investigate how a deep learning model compares with a linear statistical predictor model in terms of short-term and long-term predictive performance, and how additional engineered features, such as the ratio of uplink to downlink packets and protocol used in packet transfer, can improve the predictive performance of LSTM;
    \item within these analyses, the impact of different design parameters, including the length of training data, length of future prediction, the feature set used in machine learning, and traffic intensity, on the performance are investigated;
    \item we further extend our analysis to  the classification of the application generating the traffic, and prediction of packet and burst arrivals.  The results presented in this work pave the way for the design of traffic-aware network planning, resource management and network security.
\end{itemize}


The remainder of this paper is organized as follows:  In Section 2, we outline the related work in the area and introduce the knowledge gaps of state-of-the-art. In Section 3, we formulate the problem studied in this paper, while Section 4 presents the two methods used for solving it.  Section 5 presents the experimental evaluation results for different methods and feature sets, as wells as provides a conclusive discussion on the results. Finally, concluding remarks and future direction of research are provided in Section 6.

\section{Related work and research gap}
\label{sec:related}
We summarize state-of-the-art research on cellular traffic prediction and classification, and introduce the research gaps which motivate our work.

\smallskip
\noindent \textbf{Cellular traffic prediction.} Understanding  dynamics of cellular traffic and prediction of  future demands are, on the one hand, crucial requirements for improving resource efficiency \cite{RaUrllc}, and on the other hand, are complex problems due to the diverse set of applications that are behind the traffic. Dealing with network traffic prediction as a time series prediction, one may categorize the state-of-the-art proposed schemes into three categories: statistical learning \cite{bookrima,Seq2Seq}, machine learning \cite{LstmTrafficRaw,SpTmBigDL}, and hybrid schemes \cite{HybArimaLstm}. ARIMA and LSTM, as two popular methods of statistical learning and machine learning time series forecasting, have been compared in a variety of problems, from economics \cite{ArimaLstm,Seq2Seq,PersonDemand} to network engineering  \cite{NeuTM}. A comprehensive survey on cellular traffic prediction schemes, including convolutional and recurrent neural networks, could be found in \cite{DlForecast,EntropyCellular}. A deep learning-powered approach for prediction of overall network demand in each region of cities has been proposed in \cite{NetDemand}. In \cite{SpTmTraffic,SpTmBigDL}, the spatial and temporal correlations of the cellular traffic in different time periods and neighbouring cells, respectively, have been explored using neural networks in order to improve the accuracy of traffic prediction.  In \cite{ModelLstmDnn}, convolutional and recurrent neural networks have been combined in order to further capture dynamics of time series, and enhance the prediction performance. In \cite{NeuTM,LstmTrafficRaw}, preliminary results on network traffic prediction using LSTM have been presented, where the set of features used in the experiment and other technical details are missing. Reviewing the state-of-the-art, one observes there is a lack of research of leveraging advanced learning tools for cellular traffic prediction, selection of adequate features, especially when it comes to each user with specific set applications and  behaviours.

\smallskip
\noindent \textbf{Cellular traffic classification.}
Traffic classification has been a hot topic in computer/communication networks for more than two decades due to its vastly diverse applications in resource provisioning, billing and service prioritization, and security and anomaly detection \cite{deepcl,cls2006}.  While different  statistical  and machine learning tools have been used till now for traffic classification, e.g. refer to  \cite{lopez} and references herein, most of these works are dependent upon features which are either not available in encrypted traffic, or cannot be extracted in real time, e.g. port number and payload data \cite{lopez,deepcl}. In \cite{RnnClass}, classification of traffic using convolutional neural network using 1400 packet-based features as well as network flow features has been investigated for classification of encrypted traffic, which is too complex for a cellular network to be used for each user. Reviewing the state-of-the-art reveals that there is a need for investigation of low-complex scalable cellular traffic classification schemes (i) without looking into the packets, due to encryption and latency, (ii) without analyzing the inter-packet arrival for all packets, due to latency and complexity, and (iii) with as few numbers of features as possible. This research gap is addressed in this work.
\begin{figure}[!htb]
\vspace{-3mm}
        \centering
                \includegraphics[width=0.85\columnwidth]{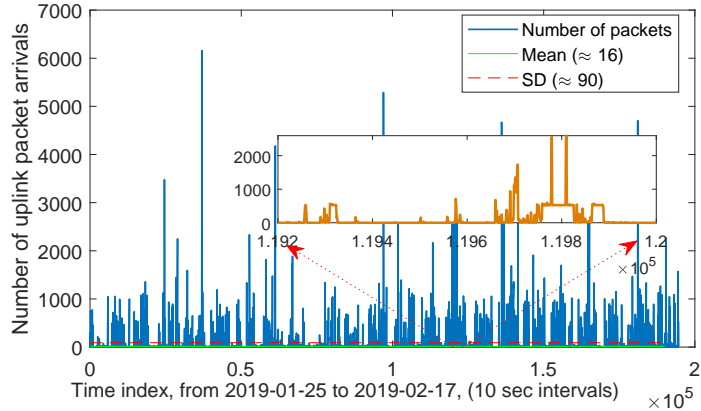}
                \vspace{-4mm}
                \caption{ The number of uplink packet arrivals  for 24 days in 10-seconds intervals} 
                \label{arr}
\vspace{-3mm}
\end{figure} 

\section{Problem description and  traffic prediction framework} 
In this section, we first provide our problem setup and formulate the research problem addressed in the paper. Then, we present the overall structure of the traffic prediction framework, which is introduced in this work. 

Consider a cellular device, on which a set of applications, denoted by $\textbf{A}$, are running, e.g., User-1 in Fig. \ref{sys}. 
At a given time interval $[t,t+\tau]$ of length $\tau$, each application could be in an \emph{active} or \emph{background} mode, based on the user behaviour. Without decoding the packets, we can define a set of features for aggregated cellular traffic in  $[t,t+\tau]$ for a specific user, such as the overall number of uplink/downlink packets and the overall size of uplink/downlink packets. Let vector $\textbf x(t)$ denote the set of features describing  the traffic in interval $[t,t+\tau]$. Furthermore, let $\textbf{X}_m(t)$ be a matrix containing the latest $m$ feature vectors of traffic for $m\ge 0$. For example, $\textbf{X}_2(t)=[\textbf{x}(t-1),\textbf{x}(t)]$. Further, denote by $\textbf s$ an indicator vector, with elements either 0 or 1. Then, given a matrix $\textbf{X}_m(t)$ and a binary indicator vector $\textbf s$, we define $\textbf{X}^s_m(t)$ the submatrix of $\textbf{X}_m(t)$, such that all respective rows, for which $s$ indicates a zero value, are removed.  For example, let
\[
\textbf{X}_m(t)=
\begin{bmatrix}
   1 &
   2 \\
   3 &
   4
\end{bmatrix}
\text{     and      }
s=[1,0] \ .
\]
Then, $\textbf{X}^s_m(t)=[1,2]$.  

Now, the research question in Section~\ref{sec:introduction} could be rewritten as:

\begin{align} 
&\textrm{Given}\quad \textbf{X}_{m}(t-1), m\ge 0;\nonumber\\ 
&\textrm{minimize}\quad L\big({\textbf X}_{-n}^\textbf{s}(t),{\textbf Y(t)}\big)\label{opp}\\
&\textrm{subject to:}\quad n\in \mathbb{Z}, n\ge 0,\nonumber
\end{align}
where $n$ is the length of the future predictions, e.g., $m=0$ for one step prediction, $\textbf Y(t)$ is of the same size as ${\textbf X}_{-n}^\textbf{s}(t)$ and represents the predicted matrix at time $t$, while $L(\cdot)$ is the desired error function, e.g., it may compute the mean squared error  between ${\textbf X}_{-n}^\textbf{s}(t)$ and $\textbf Y(t)$ when all features are scalars. 
 
\section{Time Series Prediction} \label{sec:methods}
In this section, we give a short description of the two methods benchmarked in this paper to be used within the proposed prediction framework in Section~\ref{sec:prop}. 

\subsection{The proposed traffic prediction framework} \label{sec:prop}
\begin{figure}[!htb]
\vspace{-6mm}
        \centering
                \includegraphics[width=0.95\columnwidth]{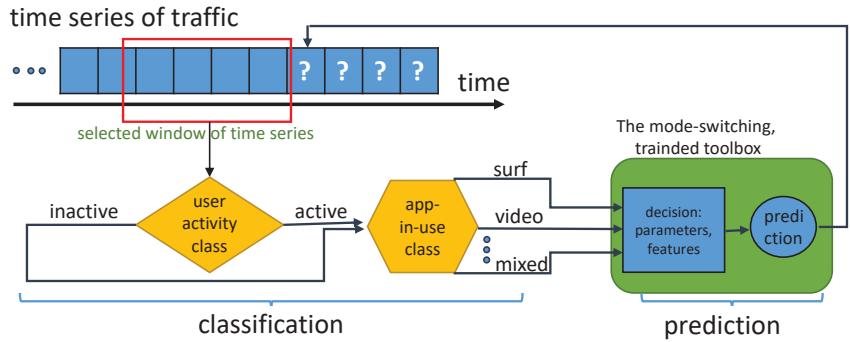}
                \caption{The proposed framework for cellular traffic prediction} 
                \label{pr}
\vspace{-10mm}
\end{figure}

Recall the challenges described in the previous section on the prediction of cellular traffic, where the major challenge consists of dependency of traffic arrival to user behavior and type of the application(s) generating the traffic. Then, as part of the solution to this problem, one may first predict the application(s) in use and behaviour of the user, and then use them as extra features in the solution. This approach for solving \eqref{opp} has been  illustrated in Fig.~\ref{pr}.  In order to realize such a framework, it is of crucial importance to first evaluate the traffic predictability and classification using only statistics of traffic with granularity $\tau$, and then, to investigate hybrid models for augmenting predictors by online classifications, and finally to investigate  traffic-aware network management design. 

\subsection{Statistical learning: ARIMA} \label{sec:arima}
The first method we consider in our work is Autoregressive integrated moving average (ARIMA), which is essentially a statistical regression model.  The predictions performed by ARIMA are based on considering the lagged values of a given time series, while at the same time accommodating non-stationarity.  ARIMA is one of the most popular linear models in statistical learning for time series forecasting, originating from three models: the autoregressive (AR) model, the moving average (MA) model, and their combination, ARMA \cite{box1976time}. 

More concretely, let $\mathcal{X} = X_1, \ldots, X_n$ define a uni-variate time series, with $X_i\in \mathbb{R}$, for each $i \in [1,n]$. A $p$-\emph{order} AR model, $AR(p)$, is defined as follows: 
\begin{equation}
X_t = c + \alpha_1 X_{t-1} + \alpha_2 X_{t-2} + \ldots + \alpha_p X_{t-p} + \epsilon_t \ ,
\end{equation}
where $X_t$ is the predicted value at time $t$, $c$ is a constant, $\alpha_1, \ldots, \alpha_p$ are the parameters of the model and $\epsilon_t$ corresponds to a white noise variable.

In a similar, a $q$-order moving average process, $MA(q)$, expresses the time series as a linear combination of its current and $q$ previous values:
\begin{equation}
X_t = \mu +  \epsilon_t + \beta_1\epsilon_{t-1} + \beta_2\epsilon_{t-2} + \ldots + \beta_q\epsilon_{t-q} \ ,
\end{equation}
where $\mu$ is the mean of $X$, $\beta_1, \ldots, \beta_q$ are the model parameters and $\epsilon_i$ corresponds to a white noise random variable.

The combination of an $AR$ and an $MA$ process coupled with their corresponding $p$ and $q$ order parameters, respectively, defines an ARMA process, denoted as $ARMA(p, q)$, and defined as follows:
\begin{equation}
X_t = AR(p) + MA(q) \ .
\end{equation}
The original limitation of ARMA is that, by definition, it can only be applied to stationary time series.  Nonetheless, non-stationary time series can be stationarized using the $d^{th}$ differentiation process, where the main objective is to eliminate any trends and seasonality, hence stabilizing the mean of the time series. This process is simply executed by computing pairwise differences between consecutive observations. For example, a first-order differentiation is defined as $X_t^{(1)} = X_t - X_{t-1}$, and a second order differentiation is defined as $X_t^{(2)} = X_t^{(1)} - X_{t-1}^{(1)}$. 

Finally, an ARIMA model, $ARIMA(p, d, q)$, is defined by three parameters $p,d,q$ \cite{mills1991time}, where $p$ and $q$ correspond to the AR and MA processes, respectively, while $d$ is the number of differentiations  performed to the original time series values, that is $X_t$ is converted to $X_t^{(d)} =\nabla^d X_t$, with $X_t^{(d)}$ being the time series value at time $t$, with differentiation applied $d$ times. The full $ARIMA(p, d, q)$ model is computed as follows:
\begin{eqnarray}
 X_t^{(d)} =& \alpha_1 X_{t-1}^{(d)} + \alpha_2 X_{t-2}^{(d)} + \ldots + \alpha_p X_{t-p}^{(d)} + \epsilon_t + c +\nonumber\\
 &\beta_1\epsilon_{t-1} + \beta_2\epsilon_{t-2} + \ldots + \beta_q\epsilon_{t-q} + \mu \ .
 \label{eq:arima}
\end{eqnarray}

\begin{figure}[!htb]
        \centering
                \includegraphics[width=0.75\columnwidth]{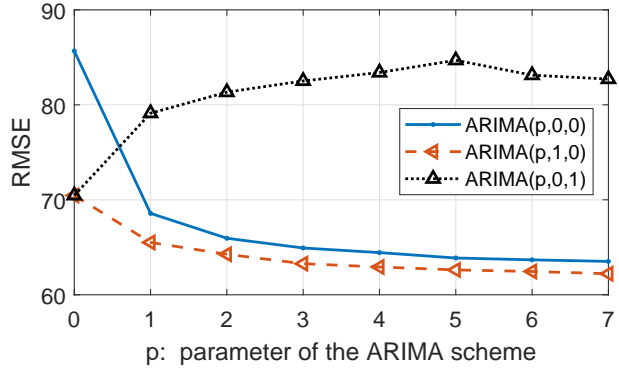}
                \caption{The RMSE performance of ARIMA($p,d,q$) for different $p,d,q$ values. } 
                \label{arimaperf}
\vspace{-4mm}
\end{figure}

\smallskip
\noindent \textbf{Finding optimized parameters.}
In this  study, the ARIMA parameters, including $p$, $d$, and $q$, are optimized by carrying out a grid search over potential values in order to locate the best set of parameters. Fig.~\ref{arimaperf} represents the root mean square error (RMSE) results for different ARIMA ($p,d,q$) configurations, when they are applied to the dataset for prediction of the number of future packet arrivals . One observes, among the presented configurations, the optimal performance is achieved by ARIMA(6,1,0) and ARIMA(7,1,0). 

\subsection{Machine learning: LSTM} \label{sec:lstm}
Next, we consider is a long short-term memory (LSTM) architecture based on a Recurrent Neural Network (RNN), a generalization of the feed forward network model for dealing with sequential data, with the addition of an ongoing internal state serving as a memory buffer for processing sequences. Let $\{X_1, \ldots, X_n\}$ define the input (features) of the RNN, $\{Y_1, \ldots, Y_n\}$ be the set of outputs, and let $\{Y_1', \ldots, Y_n'\}$ denote the actual time series observations that we aim to predict. For this study the internal state of the network is processed by Gated Recurrent Units (GRU) \cite{DBLP:journals/corr/ChoMGBSB14} defined by iterating the following three equations:
\begin{eqnarray}
r_j &=& sigm([W_rX]_j + [U_rh_{t-1}]_j) \ ,\\
z_j &=& sigm([W_zX]_j + [U_zh_{t-1}]_j)) \ ,\\
h_j^t &=& z_jh_j^{t-1} + (1-z_j)h_{new} \ ,\\
h_{new}^t &=& tanh([WX]_j + [U(r \circ h_{t-1})]_j) \ .
\end{eqnarray}
with
\begin{itemize}
    \item $r_j$: a reset gate showing if a previous state is ignored for the $j^{th}$ hidden unit,
    \item $h_{t-1}$; the previous hidden internal state $h_{t-1}$,
    \item $W$ and $U$: parameter matrices containing weights to be learned by the network,
    \item $z_j$: an update gate that determines if a hidden state should be updated with a new state $h_{new}$,
    \item $h_j^t$: the activation function of hidden unit $h_j$,
    \item $sigm(\cdot)$: the sigmod function, and
    \item $\circ$: the Hadamard product.
\end{itemize}

Finally, the loss function we optimize is the squared error, defined for all inputs as follows:
\begin{equation}
\mathcal{L} = \sum\limits_{t=1}^n (Y_t-Y_t')^2 \ .
\end{equation}
The RNN tools leveraged in this work for traffic prediction  consist of 3 layers, including the LSTM layer, with 100 hidden elements, the fully connected (FC) layer, and the regression layer. The regression layer is substituted with the softmax layer in the classification expriments.


\section{Experimental evaluation  } \label{sec:experiments}
In this section we  investigate the performance of ARIMA and LSTM powered prediction and classification tools over a real cellular user dataset.

\subsection{Dataset   } \label{sec:dataset}
We generated our own cellular traffic dataset and made part of it available online \cite{GHAA}.
The data generation was done by leveraging a packet capture tool, e.g. WireShark, at the user side. Using these tools, packets are captured at the Internet protocol (IP) level. One must note that the cellular traffic is encrypted in layer 2, and hence, the payload of captured traffic is neither accessible nor intended for analysis. The latter is due to the fact that for the realization of a low-complexity low-latency traffic prediction/classification tool, we are interested in achieving the objectives just by looking at the traffic statistics. 
 For generating labels for part of the dataset, to be used for classification, a controlled environment at the user-side is prepared in which, we filter internet connectivity   for all applications unless a  subset of applications, e.g., Skype. Then, the traffic labels will be generated based on the different filters used at different time intervals.
In our study, we focus on \textbf{seven packet features}: i) time of packet arrival/departure, ii) packet length, iii) whether the packet is uplink or downlink, iv) the source IP address, v) the destination IP address, vi) the communication protocol, e.g., UDP, and vii) the encrypted payload, where only the first three features are derived without looking into the header of packets. We experimented with different values for the interval length parameter $\tau$, and for most of our experiments $\tau$ was set to 10 seconds. Table~\ref{feat} provides the set of features for each time interval in rows, and the subsets of features used in different feature sets (FSs).  It is straightforward to infer that $\tau$ tunes a tradeoff between complexity and reliability of the prediction. If $\tau$ tends to zero, i.e., $\tau$=1 millisecond,  one can predict traffic arrival for the next $\tau$ interval with high reliability at the cost of extra effort for keeping track of data with such a fine  granularity.  On the other hand, when $\tau$ tends to seconds or minutes, the complexity and memory needed for prediction decrease, which also results in lower predictive performance during the next intervals.

\begin{table}[!tb]
\parbox{.4\linewidth}{
\centering \caption{
Features and feature sets used in our experiments.}\label{feat}
\begin{tabular}{p{2.9 cm}p{0.35 cm}p{0.35 cm}p{0.35 cm}p{0.35 cm}p{0.35 cm}p{0.35 cm}}\\
\toprule[0.5mm]
{\it  Feature sets (FSs) } & 1 & 2 & 3  & 4  & 5 & 6\\
\midrule[0.5mm]
 Num. of UL packets  & 1 &  1&1&1&1&1\\ 
  Num. of DL packets  & 1 & 0& 0&1&1&1\\ 
 Size of UL packets& 1&0&0&0&0&0\\
  Size of DL packets& 1&0&0&0&0&0\\
UL\slash DL packets& 1& 1&0&1&0&0\\
Comm. protocol& 0& 0&0&0&0&1\\
 \bottomrule[0.5mm]
\end{tabular}
}
\hfill
\parbox{.5\linewidth}{
\centering \caption{Parameter configuration in our experiments.}\label{exp}
\begin{tabular}{p{3 cm} p{3 cm}}\\
\toprule[0.5mm]
{\it Parameters }& {\it Description}\\
\midrule[0.5mm]
Traffic type&  cellular traffic\\
Capture point&  IP layer, device side\\
Length of dataset& 48 days traffic\\
RNN  for prediction (eq. classification)& [LSTM, FC, regression(eq. softmax)]\\
Time granularity, $\tau$ &   default: 10 seconds\\
 \bottomrule[0.5mm]
\end{tabular}
}
\end{table}

\subsection{Setup   } \label{sec:setup}
 The  experimental results in the following sections are presented within 3 categories, i.e. i) prediction of number of packet arrivals in future  time intervals, ii) prediction of burst occurrence in  future intervals, and iii) classification of  applications which are generating the traffic, in order to answer the three research question raised in Section~\ref{sec:introduction}. In the first two categories, we did a comprehensive set of Monte Carlo MATLAB simulations \cite{mont},  over the data set, for different lengths of the training sets, length of future prediction, feature sets used in learning and prediction, and etc. For example, each RMSE result in Fig. \ref{rmsetrain} for each scheme has been derived by averaging  over 37 experiments. In each experiment,  each scheme is trained using a train set starting from a random point of the dataset, and then is tested over 2000 future time intervals after the training set.   For the classification performance evaluation, we have leveraged 16 labeled datasets, each containing traffic from 4 mobile applications. Then, we construct 16 tests, in each test, one dataset is used for performance evaluation.  The notation of  schemes used in the experiments, extracted from the basic ARIMA and LSTM methods described in Section~\ref{sec:methods}, is as follows: (i) AR(1), which represents predicting the traffic  based on the last observation; (ii) optimized ARIMA, in which the number of lags and coefficients of ARIMA are optimized using a grid search for RMSE minimization; and (iii) LSTM(FS-$x$), in which FS-$x$ for $x\in\{1,\cdots,6\}$ represents the feature set used in the LSTM prediction/classification tool. 
 The overall configuration of experiments could be found in Table~\ref{exp}.

\smallskip
\noindent \textbf{Reproducibility.}
All experiments could be reproduced using the dataset available at the supporting Github repository \cite{GHAA}.

\subsection{Empirical results   } \label{sec:results}

In this section, we present the prediction and classification performance results in 3 subsections, including prediction of traffic intensity in future time intervals, prediction of burst events, and classification of traffic. In the first subsection, root mean square error (RMSE) is chosen as the performance indicator, while in the last two subsections, accuracy and recall are the performance indicators. 

\begin{figure}[!htb]
\vspace{-8mm}
        \centering
                \includegraphics[width=0.65\columnwidth]{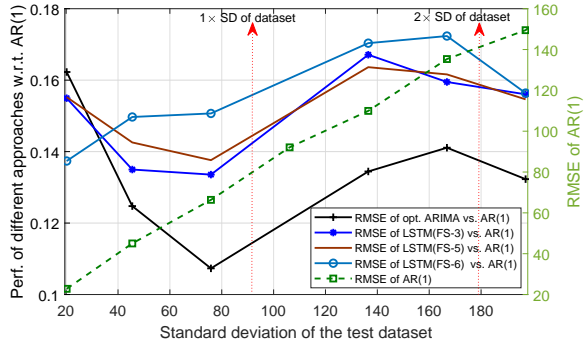}
                \caption{The RMSE performance of LSTM and ARIMA in predicting the future number of uplink packets ($\tau$=10 sec).} 
                \label{comprmse}
\vspace{-4mm}
\end{figure}

\smallskip
\noindent \textbf{Prediction of traffic intensity.} Fig. \ref{comprmse} represents the RMSE results for different ARIMA and LSTM configurations versus AR(1), when the number of uplink packets in intervals of 10 seconds is to be estimated. Towards this end, the right $y$-axis represents the absolute RMSE of AR(1) scheme,  the left $y$-axis represents the relative performance of other schemes versus AR(1), and the $x$-axis represents the standard deviation (SD) of the test dataset. The results are insightful and shed light to the  regions in which ARIMA and LSTM perform favorably, as follows. When the SD of traffic from its average value  is more than 30\% of the long-term SD of the dataset\footnote{The long-term SD of the dataset is 90.}, which is almost the case in the active mode of phone usage by human users, LSTM outperforms the benchmark schemes. On the other hand, when there is only infrequent light background traffic, which is the case on the right-end side of Fig. \ref{comprmse}, ARIMA outperforms the benchmark schemes. When we average the performance over  a 24-days dataset, we observe that  LSTM(FS-6), LSTM(FS-5), LSTM(FS-3), and optimized ARIMA outperform the AR(1) by 16\%, 14.5\%, 14\%, and 12\%, respectively, for $\tau$=10 sec.   Recall that LSTM(FS-6) keeps track of the number of uplink and  downlink packets, as well as  statistics of the communication protocol used by packets in each time interval, while LSTM(FS-5) does not care about the protocol used by packets. The superior performance of LSTM(FS-6) with regards to LSTM(FS-5), as depicted in Fig. \ref{comprmse}, represents that how adding features to the LSTM predictor can further improve the prediction performance in comparison with the linear predictors.    

\begin{figure}[!htb]
 \vspace{-8mm}
        \centering
        \includegraphics[width=0.48\columnwidth]{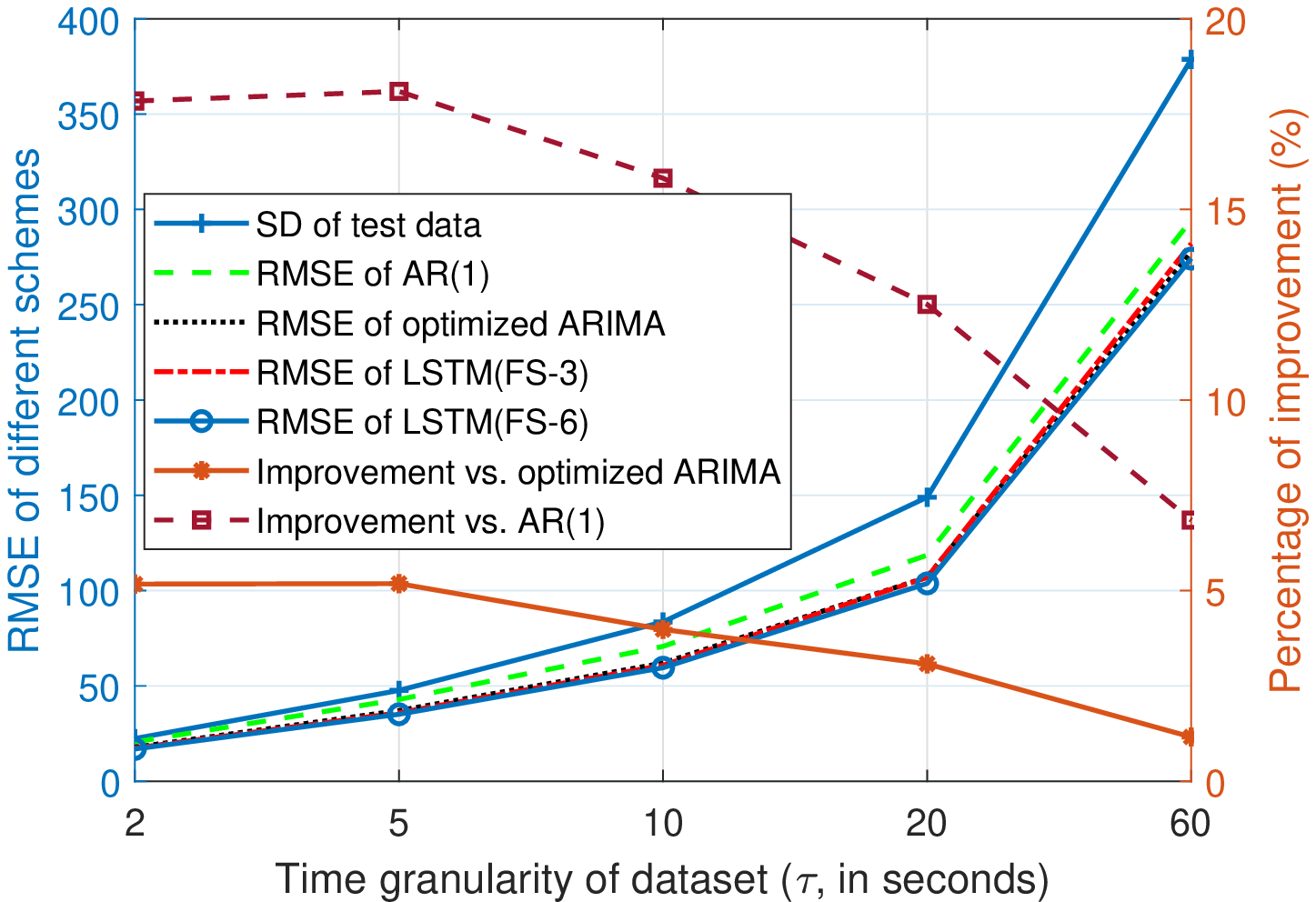}
                \includegraphics[width=0.48\columnwidth]{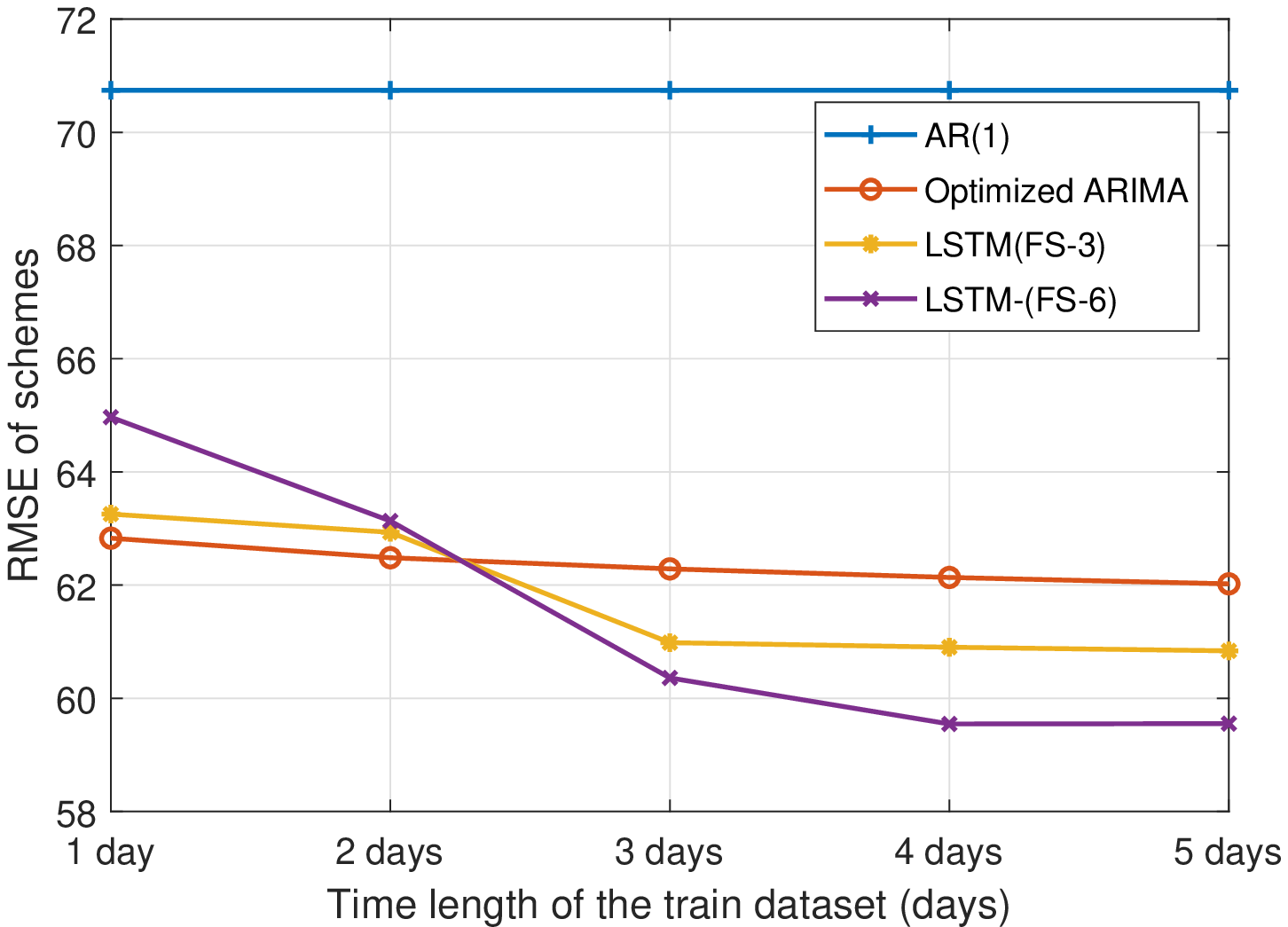}
                \caption{(left) RMSE of prediction as a function of $\tau$ (time granularity of dataset); (right) RMSE of prediction  of number of uplink packets as a function of length of the training dataset (as well as length of future prediction).} 
                \label{rmsetrain}
\vspace{-6mm}
\end{figure}

We investigate if LSTM can further outperform the benchmark schemes by increasing time-granularity of the dataset, decreasing length of future observation, and increasing length of the training set. First, let us  investigate the performance impact of $\tau$, i.e. the time granularity of dataset. Fig. \ref{rmsetrain} (left) represents the absolute (left $y$-axis) and rational (right $y$-axis) RMSE results for the proposed and benchmark schemes as a function of time granularity of dataset ($\tau$, the $x$-axis). One must further consider the fact that $\tau$ not only represents how fine we have access to the history of the traffic, but also represents the length of future prediction. It is clear that the best results for the lowest $\tau$, e.g. when $\tau=1$, the LSTM (FS-6) outperforms the optimized ARIMA by 5\% and the AR(1) by 18\%. One further observes that by increasing the $\tau$, not only the RMSE increases, but also the merits of leveraging predictors decrease, e.g. for $\tau=60$, LSTM(FS-6) outperforms AR(1) by 7\%. Now, we investigate the performance impact of length of training set on the prediction in Fig. \ref{rmsetrain} (right). One observes that the LSTM(FS-6) with poor training (1 day) even performs worse than optimized ARIMA.However, as the length of training data set increases, the RMSE performance for the LSTM predictors, especially for LSTM(FS-3) with further features,   decreases significantly. 

\begin{figure}[!htb]
 \vspace{-6mm}
        \centering
                \includegraphics[width=0.48\columnwidth]{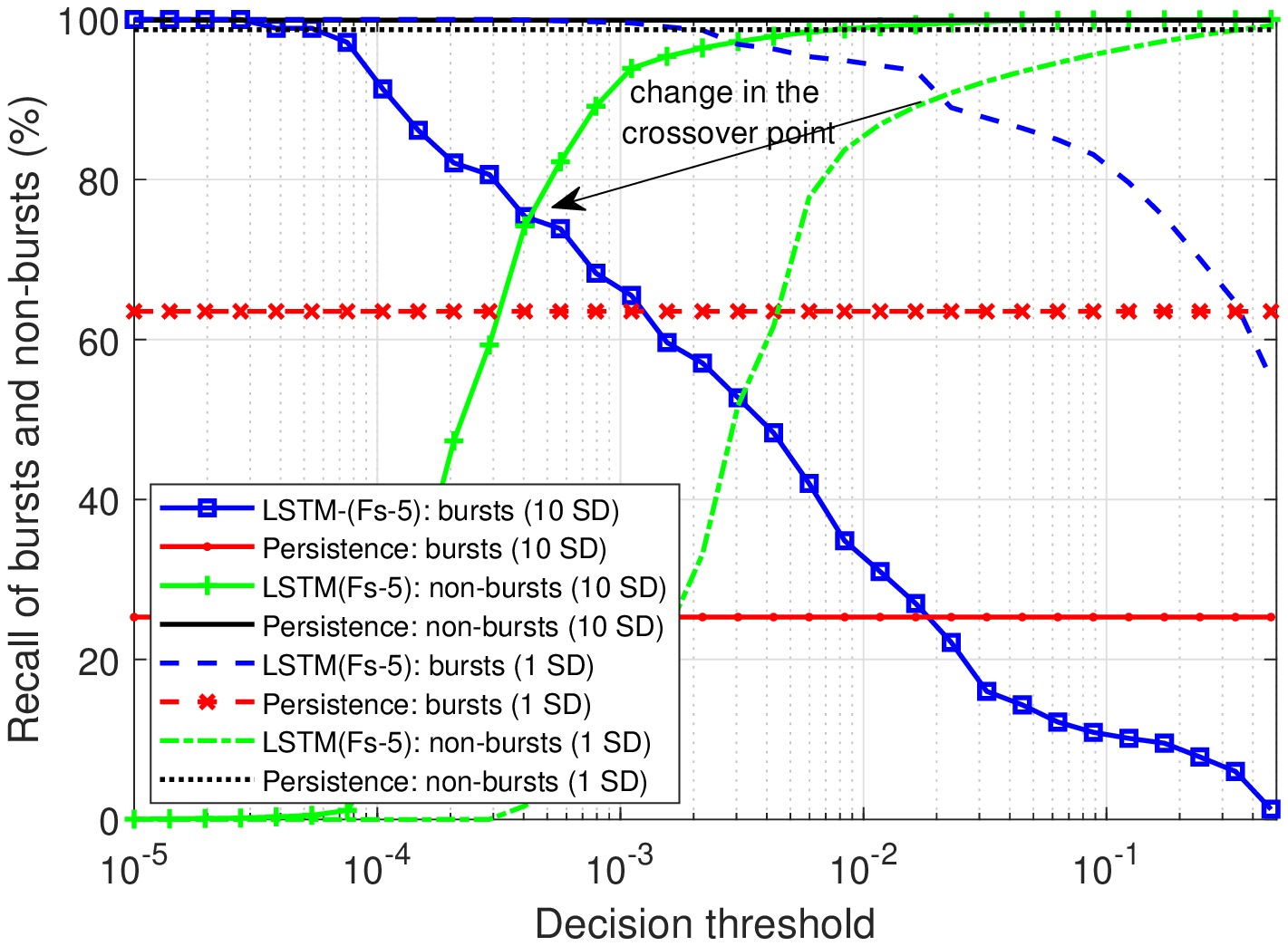}
                \includegraphics[width=0.48\columnwidth]{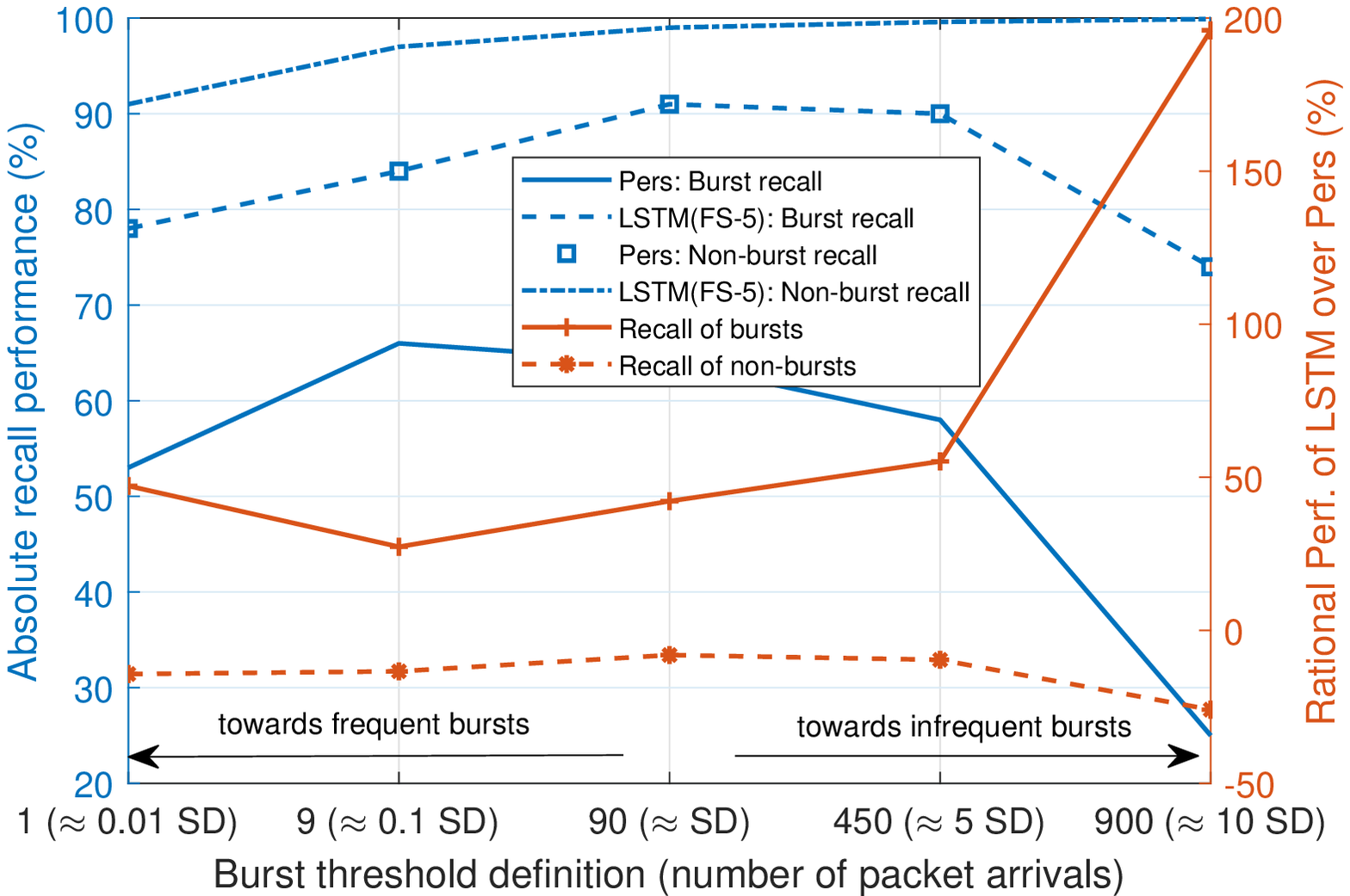}
                \caption{(left) Performance evaluation of prediction of bursts as a function of decision threshold ($\tau$=10 sec); (right) Performance evaluation of of prediction of bursts as a function of frequency of occurrence of bursts ($\tau$=10 sec).} 
                \label{tradrecalacc}
\vspace{-6mm}
\end{figure}

\smallskip
\noindent \textbf{Prediction of event bursts.}
We investigate the usefulness of the proposed schemes for burst prediction in the future time intervals. For the following experiments, we label a subset of time intervals based on the intensity of traffic, e.g. number or length  of packets, as burst. Then, based on this training dataset, we aim at predicting if a burst will happen in the next time interval or not. As a benchmark to the LSTM predictors, we compare the performance against AR(1), i.e. we estimate a time interval as burst if the previous time interval had been labeled as burst. Fig. \ref{tradrecalacc} (left) represents the recall of bursts and non-bursts for two different burst definitions. The first (second) definition treats the time intervals with more than 90 (900) uplink packet arrivals as burst, when the SD of packet arrivals in the dataset is 90. The LSTM predictor developed in this experiment returns the probability of burst occurrence in the next time interval, based on which, we need to  set a threshold probability value to declare the decision as burst or non-burst. The $x$-axis of Fig. \ref{tradrecalacc} (left) represents the decision threshold, which tunes the importance of recall and accuracy of decisions. In this figure, we observe that  the probability of missing a burst is very low in the left side, while the accuracy of decisions is low (it could be inferred from the recall of non-bursts). Furthermore, on the right side of this figure, the probability of missing bursts has been decreased, however, the accuracy of decisions has been increased. The crossover point, where the recall of bursts and non-burst match, could be an interesting point for investigating the prediction performance. In this figure for burst definition of type (1 SD), one  observes that when the decision threshold is 0.02, 91\% of burst could be predicted, while only 9\% of non-bursts are labeled as burst (false alarm).   

In Fig. \ref{tradrecalacc} (right) we observe some insightful results on the coupling between recall of predictions and degree of rareness of the bursts. In this figure, the $x$-axis represents the definition of bursts, e.g. for $x=90$, we label time intervals with more than 90 packets as  burst. From this figure, it is clear that LSTM outperforms the benchmarks in recalling the burst with a reasonable non-burst recalls cost. For example, for $x=1(\approx 0.01 SD)$, we aim at predicting if the next time interval will contain a packet or not, i.e. time intervals with a packet transmission are defined as bursts. One observes that 78\% of burst could be predicted using LSTM(FS-5), while only 28\% of non-burst are declared as bursts. Having the information that 20\% of time internals of actually burst, we infer that the accuracy of prediction has been 78\%. As the   frequency of burst occurrence decreases, i.e. we move to the right side of the figure, the recall performance of LSTM increases slightly up to some point beyond which, the recall performance starts decreasing. On the other hand, the accuracy of prediction by moving from left to right decreases significantly due to the rareness of the burst occurrence events.   The right $y$-axis represents the rational performance of LSTM versus AR(1). Clearly, LSTM outperforms AR(1) significantly, especially when bursts are occurring infrequently.

\begin{figure}[!htb]
\vspace{-8mm}
        \centering
                \includegraphics[width=0.48\columnwidth]{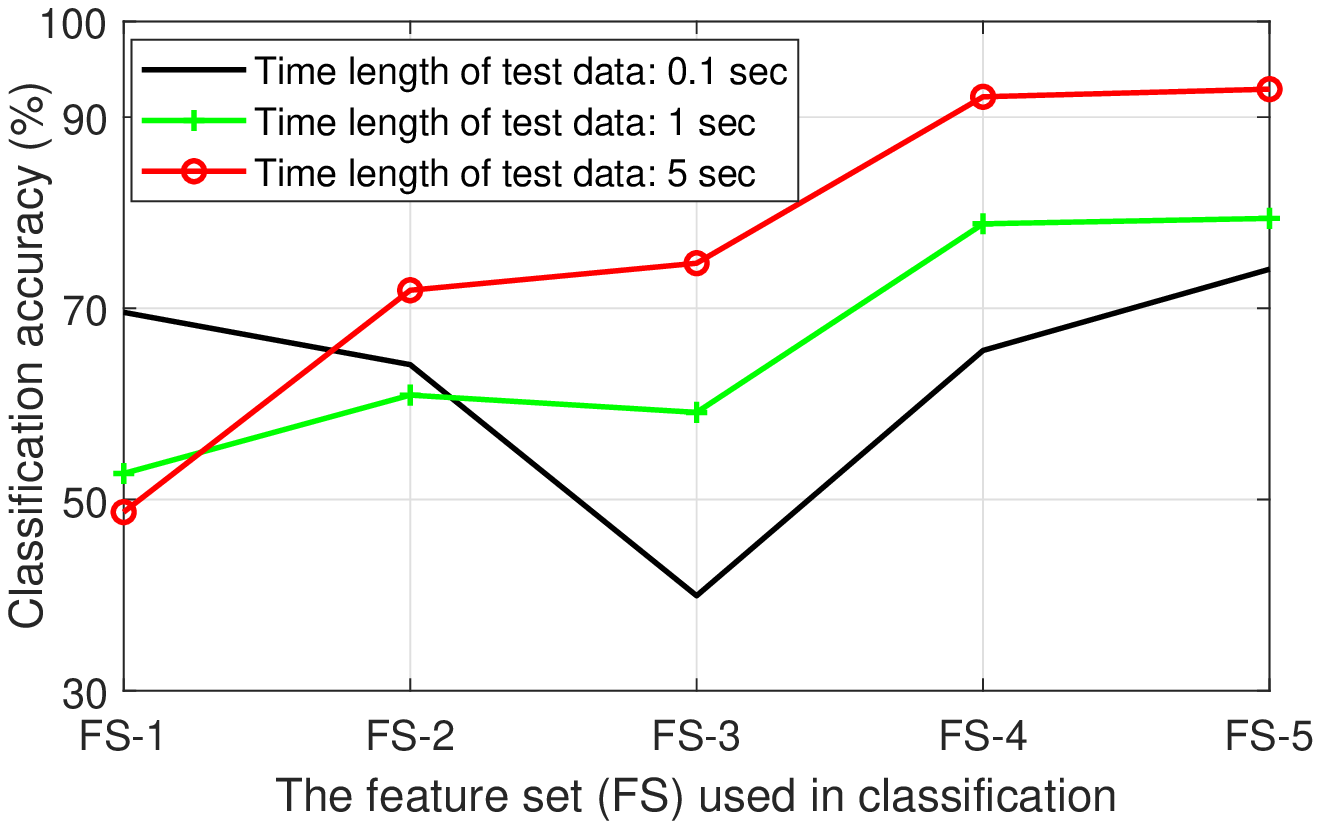}
                \includegraphics[width=0.48\columnwidth]{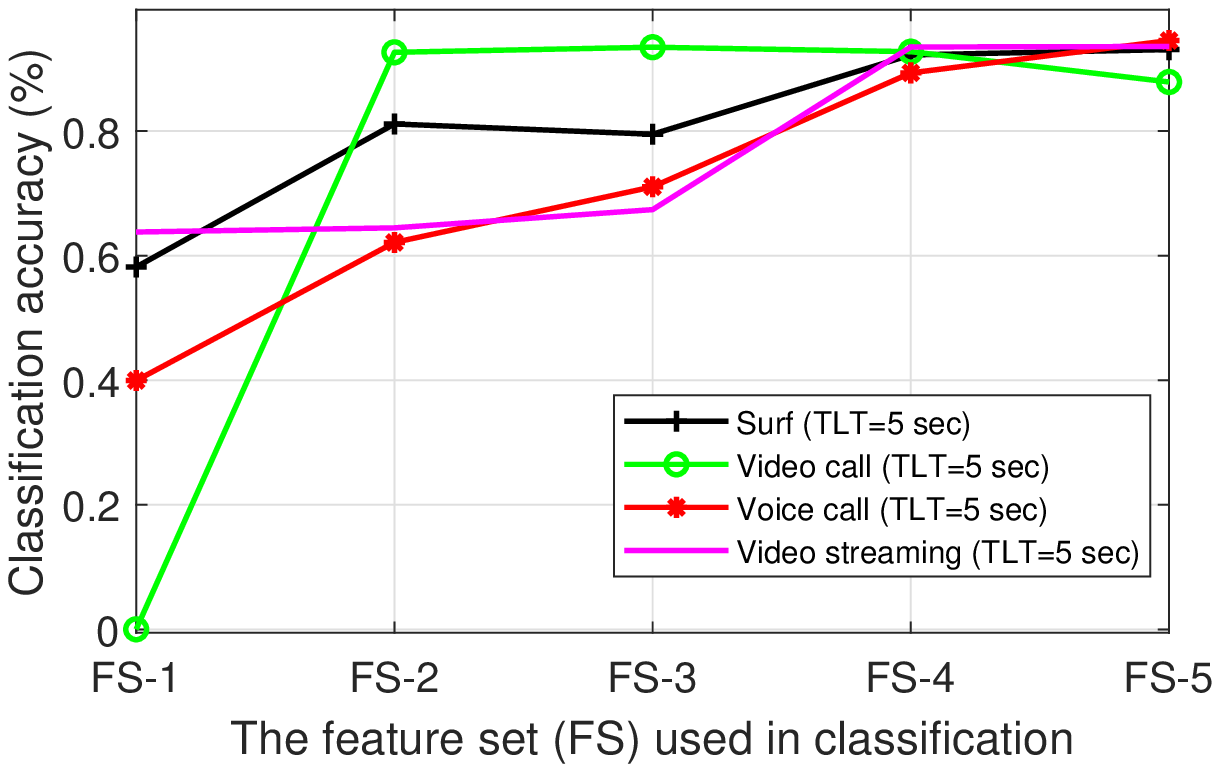}
                \caption{(left) The overall accuracy of classification as a function of the feature set used in the experiment; (right) Per application accuracy of classification as a function of the feature set used in the experiment.} 
                \label{accFs}
\vspace{-6mm}
\end{figure} 

\smallskip
\noindent \textbf{Traffic classification.}
We investigate leveraging machine learning schemes for classification of the application generating the cellular traffic in this subsection. For the classification purpose, a controlled experiment at the  user-side has been carried out in which, 4 popular applications including surfing, video calling, voice calling, and video streaming have been used by the user. Fig. \ref{accFs} (left) represents the overall accuracy of classification for different feature sets used in the machine learning tool.  One observes that the LSTM(FS-5) and LSTM(FS) outperform the others significantly in the accuracy  of classification. Furthermore, in this figure  3 curves for different lengths of the test data, to be classified, have been depicted. For example, when the length of the test data is 0.1 sec, the time granularity of dataset ($\tau$) is 0.1 sec, and  we also predict labels of intervals of length 0.1 sec. It is clear here that as the length of $\tau$ increases, the classification performance increase because we will have more evidence from the data in the test set to be matched to each class. To further observe the recall of classification for different applications, Fig. \ref{accFs} (right) represents the accuracy results per each application. One observes that the LSTM(FS-4) and LSTM(FS-5) outperform the others. It is also insightful that adding the ratio of uplink to downlink packets to FS-5, and hence constructing FS-4 (based on Table \ref{feat}), can make the prediction performance more fair for different applications. It is further insightful to observe that the choice of feature set to be used  is sensitive to the application used in the traffic dataset. In other words, FS-3, which benefits from one feature, outperforms the others in the accuracy of classification for video calling, while it results in classification error for other traffic types.

\subsection{Discussion   } \label{sec:discussion}
The experimental results represent that the accuracy of prediction strongly depends on the length of training dataset, time granularity of dataset, length of future prediction, mode of activity of the user (standard deviation of test dataset), and the feature set used in the learning scheme. The results, for example, indicate that the proposed LSTM(FS-3) is performing approximately 5\% better than optimized ARIMA, and 18\% better than AR(1) for $\tau$=10 seconds.  The results further indicated that the performance of LSTM could be further improved by designing more features related to the traffic, e.g. the protocol in use for packets, and the ratio of uplink to downlink packets.
Moreover, our experiments indicated that the design of a proper loss function, and equivalently the decision threshold,  can significantly impact the recall and accuracy performance. Furthermore, we observed that the frequency of occurrence of bursts (definition of burst), the time granularity of dataset, and length of future prediction, can also significantly impact the prediction performance. The results, for example, indicated that a busy interval, i.e. an interval with at least one packet, could be predicted by 78\% accuracy as well as recall.
The experimental results represented the facts that, first, accuracy and recall performance of classification is highly dependent on the feature set used in the classification. For example, a feature set that can achieve an accuracy of 90\% for classification of one application may result in a recall of 10\% for another application. Then, the choice of feature set should be in accordance with the set of applications used by the user.  Second, if we can tolerate delay in the decision, e.g. 5 sec, the classification performance will be much more accurate when we gather more information and decide on longer time intervals. The overall accuracy performance for different applications using the developed classification tool is approximately 90\%. 

\section{Conclusions} \label{sec:conclusions}
In this work, the feasibility of per-user traffic prediction for cellular networks has been  investigated. Towards this end, a framework for cellular traffic prediction has been introduced, which leverages statistical/ machine learning units for traffic classification and prediction. A comprehensive comparative analysis of prediction tools based on statistical learning, ARIMA, and the one based on machine learning, LSTM, has been carried out, under different traffic circumstances and design parameter selections.  The LSTM model, in particular when augmented by additional features like the ratio of uplink to downlink packets and the communication protocol used in prior packet transfers, exhibited demonstrable improvement over the ARIMA model for  future traffic predictions. Furthermore,  use fullness of the developed LSTM model for classification of cellular traffic has been investigated, where the results represent high sensitivity of accuracy and recall of classification to the feature set in use. Additional investigations could be performed regarding  making the prediction tool mode-switching, in order to reconfigure the feature set and prediction parameters based on the changes in the behaviour of user/applications in an hourly/daily basis.

 \bibliographystyle{splncs04}
 \bibliography{paperbib}
\end{document}